\documentclass[aps,pra,showpacs]{revtex4}

\usepackage{graphics}
\usepackage{natbib}
\usepackage{epsfig,latexsym}

\newcounter{saveeqn}

\begin{document}

\title{Dark resonances as a probe for the motional state of a single ion}
\author{C. Lisowski}
\author{M. Knoop} \email{Martina.Knoop@up.univ-mrs.fr}
\author{C. Champenois}
\author{G. Hagel}
\author{M. Vedel}
\author{F. Vedel }

\affiliation{Physique des Interactions Ioniques et Mol\'eculaires
(CNRS UMR 6633), Universit\'e de Provence, Centre de Saint
J\'er\^ome, Case C21, 13397 Marseille Cedex 20, France}

\date{Received: \today / Revised version: date}

\begin{abstract}
Single, rf-trapped ions find various applications ranging from
metrology to quantum computation. High-resolution interrogation of
an extremely weak transition under best observation conditions
requires an ion almost at rest. To avoid line-broadening effects
such as the second order Doppler effect or rf heating in the
absence of laser cooling, excess micromotion has to be eliminated
as far as possible. In this work  the motional state of a confined
three-level ion is probed, taking advantage of the high
sensitivity of observed dark resonances to the trapped ion's
velocity.  Excess micromotion is controlled by monitoring the dark
resonance contrast with varying laser beam geometry. The influence
of different parameters such as the cooling laser intensity has
been investigated experimentally and numerically.
\end{abstract}

\pacs{32.80.Pj, 39.30.+w}
\maketitle

\section{Introduction}

Dark resonances in a three-level system, also called coherent
population trapping,  are a well-known example of destructive
quantum interference based on the interaction of two light fields
generating the coherent superposition of two atomic states.
Various applications of the coherent population trapping scheme
can be found in atomic physics  from high-resolution spectroscopy
to sub-recoil laser-cooling of an atom cloud \cite{arimondo96} or
EIT-cooling of single ions \cite{morigi00,roos00,morigi03}. Dark
resonances have been readily observed in trapped ions, almost
exclusively with copropagating laser beams
\cite{janik85,stalgies96,reiss02,kurth95}. In general the observed
splitting of the various Zeeman levels is used for an exact
quantitative calibration of the local magnetic field.

Single ions confined in radiofrequency traps are ideal candidates
for different applications of high-resolution spectroscopy such as
quantum information or frequency standards in the optical domain
\cite{Gill2002}.  They can be stored in the trap from hours up to
months in a quasi interaction-free environment, oscillating at
fixed frequencies  in an effective pseudo-potential well. Laser
cooling of a single trapped ion allows to reach the Doppler
cooling limit which is in the mK-range. To eliminate residual
Doppler broadening on a given transition, the ion's excursion
along the laser propagation direction must be smaller than the
inverse wavenumber to fulfill the Lamb-Dicke condition
\cite{dicke53}. The excitation spectrum of the considered
transition is then decomposed in a central carrier and a few
sidebands separated by the ion's frequencies of motion (Lamb-Dicke
regime).

The excellent spatial localisation required by the Lamb-Dicke
condition can only be achieved in the low-field region of the rf
trapping field. Experimentally, this is realized with the
confinement of  single ions in the potential minimum of the trap,
with one exception where two ions have been stored in an
extraordinarily steep, elliptical potential \cite{king98}. For
this purpose, the pseudo-potential at the position of the ion has
to be nearly perfectly symmetric. Flaws in the potential due to
asymmetries or patch potentials caused by atom deposition during
ion creation, may distort the created potential and deviate the
minimum of the trap's AC electric field from the minimum of the
pseudo-potential well. The resulting displacement of the average
position of the ion in the trap causes an increase of the velocity
amplitude at the confinement frequency, called excess micromotion,
and has to be corrected. The influence of parasitic potentials can
be compensated by direct voltages applied on supplementary
correction electrodes in the vicinity of the trap.

Different experimental techniques have been employed to reduce the
excess micromotion of a single trapped ion (\cite{berkeland98} and
references therein, \cite{raab00}). In this paper we present an
additional and straightforward method for the localisation of an
ion in the miniature trap. We have used coherent population
trapping as a tool to minimize the micromotion of the confined
ion. Dark resonances depend ideally only on the lifetimes of the
two involved atomic states, which in the present experiment are
the 4$S_{1/2}$ ground state and the  metastable 3$D_{3/2}$  state
(natural lifetime close to 1 s \cite{knoop95,lidberg99}) of a
single calcium ion confined in a miniature cylindrical ring trap.
In practice, the lifetime of the dark state is shortened by the
linewidth of the lasers, leading to a reduction of the visibility
of the dark resonance. The oscillating Doppler shifts encountered
by the ion also tend to wash out the dark resonance, if the dark
state is not stable on the observation time scale. The study of
the phenomenon reveals that the maximum sensitivity to the
velocity amplitude of the ion is reached with counterpropagating
laser beams. The proposed technique is based on the fact that the
depth of dark resonances observed in the fluorescence of a single
ion increases with the degree of immobilisation of the ion. Excess
micromotion drastically reduces the lifetime of the dark state,
smoothing the observed dark resonance dip.

The first section of this article  describes the micromotion
generated in the radiofrequency trap, the existing methods to
reduce this effect are very briefly reviewed. We then introduce
the formalism  for dark resonances, along with the modelling used
to obtain  a high degree of control of the experimental
conditions. The subsequent section presents numerical results,
while the last section of this work is devoted to the experimental
observation of dark resonances and its use as a tool to reduce
excess  micromotion. The influence of the main control parameters
(laser power, beam geometry and magnetic field) is discussed.

Although we try to keep our discussion of the dark resonances in a
three-level system as general as possible, we use the case of a
single trapped calcium ion as an illustration throughout the
description of the modelling. Experimental details are given in
section \ref{exp}, the level scheme with the relevant transitions
of the calcium ion is given in figure \ref{fig:expset}.

\section{Micromotion in a Paul trap}

The motion of a trapped ion in an ideal Paul trap is  described by
the Mathieu equations \cite{paul90}:
\begin{equation}
  \frac{d^{2}u}{d\tau^2}+ [a_u - 2 q_u \, \cos (2\tau)]\,u = 0 \ \  \mathrm{with}
\ \   \tau = \frac{\Omega}{2}t \label{eq:mathieu}
\end{equation}
valid for the three directions $u=x, y, z$.
 The values of the coefficients $a_u$ and $q_u$ which determine the range of stable
confinement of a particle are defined by the trapping voltage
$V_{AC}$ and its frequency $\Omega/2\pi$, the superimposed
DC-voltage $V_{DC}$, the radius $r_0$ of the trap as well as the
$e/m$ ratio of the trapped ion. To first order in $a_u$ and $q_u$,
the stable solution of equation (\ref{eq:mathieu}) is
\begin{equation}
  u(t) =  R_u \ \cos\: \omega_u t \,(1 + \frac{q_u}{2}\  \cos \:\Omega t)
  \label{eq:solmathieu}
\end{equation}

The motion of the confined ion  is composed of the harmonic
oscillation at frequencies $\omega_u/2\pi$ with amplitude $R_u$
called "secular" motion, and the AC-driven "micromotion" at the
frequency of the trapping field $\Omega/2\pi$. In opposition to
the secular motion, micromotion cannot be cooled because it is
driven motion. According to Eq.(\ref{eq:solmathieu}), a smaller
secular motion amplitude $R_u$ leads to a reduced contribution of
the micromotion.  A displacement $D_u$ of the average position of
the ion,  due to asymmetries in the geometry of the trap and the
applied electric fields,  causes excess micromotion, and can be
taken into account like
\begin{equation}
  u(t) =  (R_u \cos\: \omega_u t + D_u)(1 + \frac{q_u}{2}  \cos\: \Omega t)
  \label{eq:excess}
\end{equation}
in the equation of motion. This excess micromotion can possibly
prevent the access to the Lamb-Dicke regime. A complete review of
origin and consequences of the micromotion is given in
\cite{berkeland98}.

Three experimental methods are commonly used to control excess
micromotion of an ion  in a radiofrequency trap. Best results are
obtained by using all of them as the collected information is
complementary in the three cases. The control parameter is in
either case the set of DC-voltages applied to the compensation
electrodes surrounding the trap. The most simple approach is the
observation of the spatial displacement of a trapped ion as the
confining potential is lowered. Parasite potentials then gain in
importance and move the ion out of the trap center. This method
requires spatial detection of the ion's fluorescence and
 is limited to the plane of observation. Another means
for the rejection of excess micromotion is to probe the
fluorescence linewidth of the laser-cooling transition. When the
blue laser is scanned, only the low-frequency side of the atomic
resonance is experimentally visible, as the high-frequency
detunings induce a Doppler laser heating effect, causing a sudden
drop in fluorescence signal. This profile is difficult to
calibrate in  terms of absolute temperature, but gives a good
relative control signal of the ion's kinetic energy when laser
powers are fixed. At a given laser frequency, the half-maximum
fluorescence signal  decreases for better compensation, as the ion
becomes colder and its blue transition linewidth narrower. A more
specific observation of the micromotion's influence is the measure
of the correlation of the emitted fluorescence photons with the
confinement frequency $\Omega/2\pi$ \cite{berkeland98}. Actually,
the oscillatory motion of the ion causes a frequency modulation of
the laser field in the rest frame of the ion, and induces a
modulation of the emitted photons at the trapping frequency
$\Omega/2\pi$. To cancel the micromotion,  the amplitude of this
modulation signal has to be minimized. Best compensation results
are obtained by the use of different laser axes for an access  to
the different projections in space.

The ultimate measurement to determine the absolute degree of
localisation of the ion in the trap is the detection of
vibrational sidebands in the  spectrum of the ion. The trapped ion
oscillates at secular frequencies $\omega_u/2 \pi$ and at the
confinement frequency $\Omega/2 \pi$. The oscillatory motion at a
fixed frequency creates a frequency modulation of the laser
excitation through the  Doppler effect and leads to sidebands in
the ion spectrum. The relative height of the sidebands is related
to the amplitude of the corresponding oscillation, and their
minimization is thus an indicator for the localization of the ion
\cite{diedrich89}. One experimental realisation consists in
monitoring sidebands in the emission spectrum of a single ion
using a heterodyne technique \cite{raab00}. However, in the
majority of cases, sideband observation is made in the excitation
spectrum of the ion. This  requires that the natural linewidth of
the excited transition is inferior to the ion's vibrational
frequencies. In the present experiment, this must be realized on
the ultra-narrow 4$S_{1/2}$ - 3$D_{5/2}$ transition using the
quantum jump technique. The experimental realization necessitates
a highly stabilized laser to excite the 4$S - 3D$ transition and
the need to work on quantum jump statistics for the detection of
this dipole-forbidden transition.

In the following we show how the visibility of the dark resonance
varies with experimental conditions and how we can use
 this observation to  probe  the minimisation
of the micromotion of the confined particle.

\section{Formalism and modelling}
In a three-level system driven by two lasers in  $\Lambda$
configuration,  coherent superposition of the atomic states
coupled by the radiation fields  leads to the appearance of a
non-absorbing state, when the detunings of the two lasers are
equal. In this case, one of the stationary states $\left |
\psi_{NC} \right>$  is a coherent superposition of the ground and
metastable states $S_{1/2}$ and $D_{3/2}$, which is not coupled to
the excited state $P_{1/2}$ by the atom-laser interaction $V_{L}$
($\left<{\rm P}_{1/2}\right| V_{L} \left| \psi_{NC}\right>=0$).
Once in this state, the atom can neither absorb nor scatter
photons and the fluorescence disappears. This feature, called dark
resonance, has been used to cool atoms below the recoil limit by
velocity selective coherent population trapping \cite{aspect88},
and ions to the vibrational ground state by electromagnetically
induced transparency  \cite{roos00}.

In the general case,  the two transitions are driven by different
lasers, the non-coupled state then depends on the relative phase
of the lasers,  labelled (B) and (R) for blue and red,  like:
\begin{equation} \left | \psi_{NC}
\right>=e^{-i(\omega_Bt+\phi_B)}\frac{-\Omega_B\left |{\rm
D}_{3/2}
\right>+\Omega_Re^{-i((\omega_R-\omega_B)t+\phi_R-\phi_B)}\left |
{\rm S}_{1/2} \right>}{\overline{\Omega}} \end{equation} with
$\overline{\Omega}=\sqrt{\Omega_B^2+\Omega_R^2}$, where we suppose
the  two Rabi frequencies  $\Omega_B$ and $\Omega_R$, to be real.
This dependance implies a high stability of the applied lasers to
observe the complete extinction of the emitted fluorescence.

If the travelling wave nature of the laser fields and the motion
of the atom are taken into account, the laser couples $ \left |
{\rm P}_{1/2}, p \right>$ with  $ \left | {\rm S}_{1/2}, p-\hbar
k_B \right>$ and $ \left | {\rm D}_{3/2}, p-\hbar k _R \right>$,
where $p$, $\hbar k_B$ and $\hbar k_R$ are respectively the
projection of the atom momentum, the blue and the red photon
momentum, along the common propagation axis of the two lasers.
Then, the non coupled state is not an eigenstate of the kinetic
energy and is coupled to its orthogonal state  $ \left | \psi_C
\right>$ by the kinetic energy operator:
\begin{equation} \left | \psi_{C}
\right>=\frac{\Omega_Re^{-i(\omega_Rt+\phi_R)}\left |{\rm
D}_{3/2}, p-\hbar k_R
\right>+\Omega_Be^{-i(\omega_Bt+\phi_B)}\left | {\rm S}_{1/2},
p-\hbar k_B \right>}{\overline{\Omega}} \end{equation} In the
basis of the atom dressed by $N_B$ blue photons and $N_R$ red
photons, we study the evolution of the system inside the family
state $\mathcal{F}_p$ :
\begin{equation}  \mathcal{F}_p = \left\{ \left | S_{1/2}, p-\hbar k_B, N_B+1, N_R  \right> ; \left| P_{1/2}, p, N_B, N_R  \right> ; \left
|{\rm D}_{3/2}, p-\hbar k_R, N_B, N_R+1 \right> \right\}
\end{equation}
The lifetime of the non-coupled state is shortened by its coupling
to its orthogonal  state which is
\begin{equation}\label{eq:nc} \left<\psi_C\right| H_0+\frac{p^2}{2m}
\left|
\psi_{NC}\right>=\frac{\Omega_B\Omega_R}{\overline{\Omega}^2}
\left(\hbar(\Delta_B^{'}-\Delta_R^{'})+p\frac{\hbar(k_R-k_B)}{m}\right)
\end{equation}
where $\Delta_R^{'}$ and $\Delta_B^{'}$ are the laser detunings
corrected by the photon recoil energy:
$\Delta^{'}=\omega-\omega_{at}+\hbar^2k^2/2m$. In the case of two
identical-wavelength transitions  driven by copropagating laser
beams ($k_R=k_B$), the non coupled state is  stationary as soon as
the corrected detunings are equal. If the two laser beams are
counterpropagating or in the more general case of different
wavelengths ($|k_R|\neq|k_B|$), the non-coupled state is
stationary only if the atom is at rest in the excited state
($p=0$).  $ \left(-\Omega_B\left |{\rm D}_{3/2}, -\hbar k_R
\right>+\Omega_R\left | {\rm S}_{1/2}, -\hbar k_B
\right>\right)/\;\overline{\Omega}$ is then a perfect trap state,
as long as we neglect its finite lifetime caused by spontaneous
emission. In the case of a moving atom ($p \neq 0$), the dark
resonance condition can be interpreted as an equality of the two
detunings corrected by the Doppler shifts :
\begin{equation}\label{eq:doppler}
\Delta_B^{'}-\Delta_R^{'}-(k_B-k_R)\;p/m=0 \end{equation}

It appears from  equations (\ref{eq:nc}) and (\ref{eq:doppler})
that the lifetime of the non coupled (or dark) state is reduced by
the oscillating Doppler shifts as the relative laser detuning is
fixed. The observation of the dark resonances can then be used to
quantify the motional state of an ion, and the highest sensitivity
to this motion is obtained for $k_R\;=\; -\;k_B$.

To identify the dependance of the dark resonance profile on each
experimental parameter (laser linewidth, laser intensity and
detuning, motional state of the ion), we numerically studied the
atomic system through the evolution of its density matrix. As the
quantity of detected photons is proportional to the population of
the 4$P_{1/2}$-level, the dark resonances are visualized by
calculating this population  as a function of the red laser
detuning. The states involved in the simulation are the eight
Zeeman sublevels of the  $S_{1/2}, P_{1/2}, D_{3/2}$ states,
coupled by the blue laser (B) on the 4$S_{1/2}\rightarrow
4P_{1/2}$ transition and the red laser (R) on the
3$D_{3/2}\rightarrow 4P_{1/2}$ transition. The blue and red laser
linewidths $\Gamma_{LB}$ and $\Gamma_{LR}$ (FWHM) are taken into
account by relaxation on coherences. The mechanical effect of
light on the motion of the atom is neglected as well as the recoil
energy which is much smaller than the relevant energy scale of our
simulations. The motion of the ion is then treated classically and
taken into account by a time-dependent Doppler shift, using the
same approach as in \cite{schubert89} :
\begin{equation}\label{eq:doppler_2}
\Delta_B(t)=\Delta^0_B-k_BV(t) \ ; \ \Delta_R(t)=\Delta^0_R\pm
k_RV(t)
\end{equation}
where the sign of the red-transition Doppler shift depends on the
propagation configuration  of the lasers ($-$ for copropagating
beams, $+$ for counterpropagating beams). $ \Delta^0$ is the
detuning set by the laser frequency. For the sake of simplicity,
we suppose here that only one secular frequency ($\omega_r$)
contributes to the Doppler shift, which can be written, according
to the equation of motion (\ref{eq:excess}) with $V_0=\omega_r
R_r$:
\begin{equation}\label{eq:doppler_osc}
\Delta(t)=\Delta^0 \pm  kV_0\left(\sin\omega_r
t\left(1+\frac{q_r}{2}\cos\Omega t\right)+\sqrt{2}\cos\omega_r
t\sin\Omega t + \frac{D_r}{R_r}\sqrt{2}\sin\Omega t  \right)
\end{equation}

using the relation  $\omega_r=q_r\Omega/(2\sqrt{2})$ valid in the
adiabatic approximation \cite{dehmelt67} for a trap with zero DC
field applied ($a_r=0$).

The natural widths of the involved atomic transitions are of the
same order as the secular and rf trapping frequency
($\Gamma_B/2\pi=22.5$~MHz, $\Gamma_R/2\pi = 1.35$~MHz,
$\omega_r/2\pi = 1$~MHz, $\Omega/2\pi = 11.6$~MHz).  For the
modelling of the internal state the evolution of the density
matrix has to be followed and integrated over several periods,
until a steady state is reached. Technically convergence is
accomplished when the level populations averaged over one secular
period are not modified by adding  one period to the integration
time.

If the magnetic field is set to zero, the atomic population is
trapped in the $D_{3/2}, m_J=\pm 3/2$ states and fluorescence
disappears and hides the dark resonance we are looking for. To
prevent this optical pumping and keep the detected fluorescence
signal as high as possible, a minimal magnetic field must be
applied (see section \ref{sec:hanle} for details).

\section{Numerical results}
\label{result_simul} Numerical simulations using the described
model have been carried out to evidence the influence of the
excess micromotion, the laser intensity and the laser beam
geometry. To describe the dark resonance behavior in a
quantitative way, we define the "contrast" $\mathcal{C}$ of the
dark resonance   as the depth of the observed dip divided by the
total signal we would observe at the same frequency if there was
no dark resonance. This is done by numerical extrapolation of the
experimental signal.

\begin{figure}
\begin{center} \resizebox{0.75\textwidth}{!}{\epsfig{file=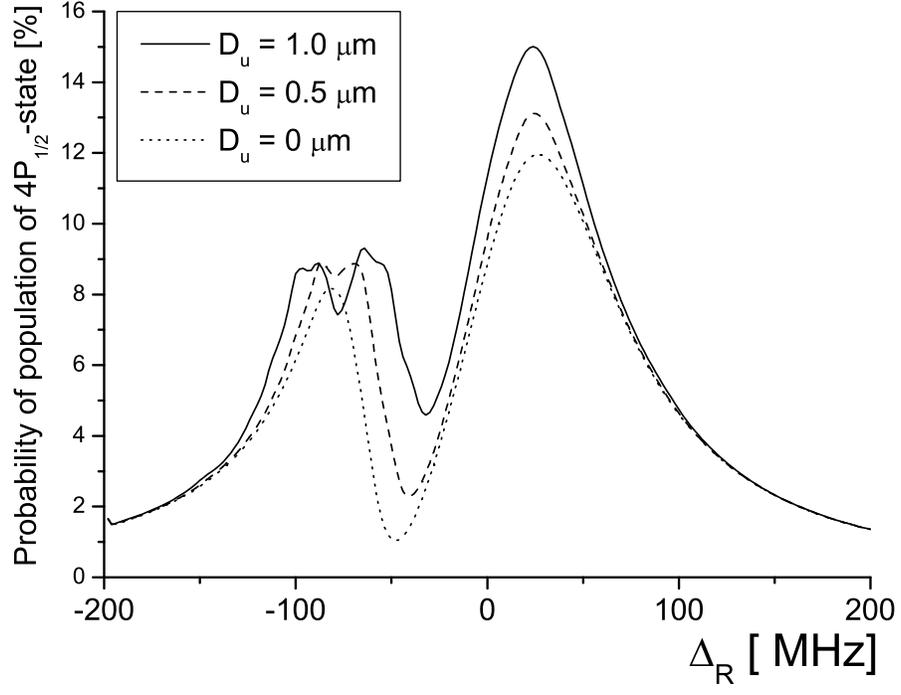}}
\end{center}
\caption{Numerical simulation of dark resonances in the
counter-propagating laser beam geometry, for a velocity amplitude
$V_0=1$ m/s and  different displacements $D_u$ off the trap
center. ($\Omega_B$ = 120~MHz, $\Omega_R$ = 30~MHz, $\Delta_B$ =
-50~MHz.)} \label{fig:simul}
\end{figure}

The sensitivity of the dark resonances to the excess micromotion
of the ion is illustrated in figure \ref{fig:simul} where the
probability of occupation of the P$_{1/2}$ state is plotted versus
the detuning of the red laser with fixed amplitude  of secular
oscillation $V_0$ and for identical laser parameters. To keep the
discussion general and take advantage of the described method even
for a non-ideal system, the velocity amplitude $V_0$ is chosen to
be 1 m/s which corresponds to a temperature of ten times the
theoretical Doppler laser cooling limit for calcium ions. Anyway,
simulations show that for the chosen laser parameters, the dark
resonance can be observed up to temperatures a hundred times the
Doppler cooling limit. The laser linewidths used in the simulation
($\Gamma_{LB}=1$~MHz, $\Gamma_{LR}=0.2$~MHz) are the measured
linewidths for the lasers in our experiment. The Rabi frequencies
($\Omega_B=120$~MHz, $\Omega_R=30$~MHz) and the blue laser
detuning ($\Delta_B=-50$~MHz) are the highest values chosen such
that the simulations reproduce the  width and profile of the
experimental spectra. The precision of this fit is of the order of
10\%, which is sufficient for  the comparison we want to make. As
in our experiment, the magnetic field has been set to 1 Gauss.
With the laser linewidths considered here, the splitting of the
Zeeman sublevels can not be observed in the dark resonance.

The spectra obtained for three different  displacements $D_u$ in
the counter-propagating laser beam configuration are plotted in
figure \ref{fig:simul}. The depth of the dark resonance is reduced
for increasing displacement and illustrates how the observation of
the contrast of dark resonances can be used for the reduction of
the excess micromotion. The curve with zero excess micromotion
($D_u = 0$) shows a contrast $\mathcal{C}$ of 91\%. Finite laser
linewidths and  secular motion prevent the dark resonance dip to
fall completely to the background light level and limit the
maximum contrast.

When the displacement $D_u$ increases from 0 to 0.5 and 1$\mu$m,
the contrast is reduced to 83\% and 73\%, respectively.  With the
laser parameters given above, the blue spectral linewidths
corresponding to the shown spectra do not increase, and the slope
of the blue resonance transition is unchanged, as the power
broadening is too important. This means that neither the rf
correlation technique nor the measurement of the spectral
linewidth can give information about the reduction of the
micromotion, whereas the contrast of the dark resonance can vary
from 73\% to 91\%, which can be detected experimentally with high
efficiency.

\begin{figure}
\begin{center}
   \resizebox{0.75\textwidth}{!}{ \epsfig{file=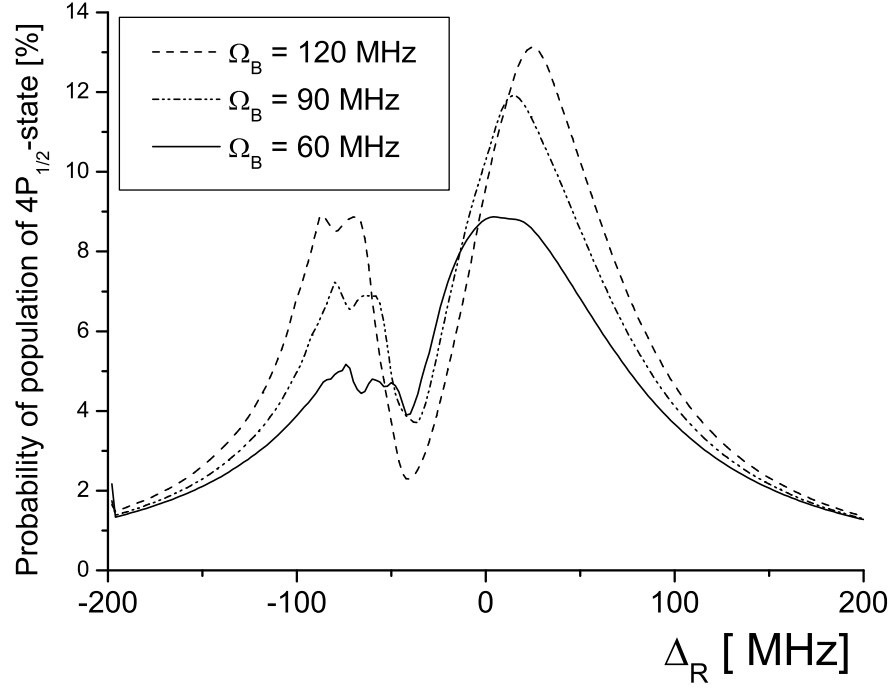}}
\end{center}
\caption{Numerical simulation of the observable fluorescence
signal for different blue Rabi frequencies $\Omega_B$.  The two
laser beams are counterpropagating and the motion of the ion is
described by $V_0$ = 1 m/s and $D_u$ = 0.5~$\mu$m. Micromotion
sidebands are visible in the spectrum at lower $\Omega_B$.
($\Omega_R$ = 30~MHz, $\Delta_B$ = -50~MHz.) }
\label{fig:simul_rabi}
\end{figure}

Observation of dark resonances can be made easier by using a
higher Rabi frequency $\Omega_B$ for the blue laser while keeping
$\Omega_R$ fixed, even though the dark resonance phenomenon occurs
independently of the employed Rabi frequencies. In the dressed
atom picture, higher Rabi frequencies increase the splitting of
the   states resulting from the coupling of $S_{1/2}$ and
$P_{1/2}$ by the blue laser (see inset of figure \ref{fig:expset}
for details). The spectrum observed by scanning the red laser is
then split into the Autler-Townes doublet and the sharp dark
resonance profile is replaced by a broader feature which is less
sensitive to the oscillating Doppler shifts. The influence of
different $\Omega_B$ is illustrated in figure \ref{fig:simul_rabi}
where the excitation probability to $P_{1/2}$ is plotted for
different Rabi frequencies. The graphs show that the increased
doublet splitting can also be advantageous for the reduction of
the excess micromotion, as $\mathcal{C}$ has a minimum value,
depending on the laser parameters $\Omega_B$, $\Omega_R$, and
$\Delta_B$, even if the dark resonance is washed out. This can be
used in particular to start the compensation process while the
excess micromotion amplitude is still high.
 In the cases
studied in figure \ref{fig:simul_rabi}, the contrast is 53\% for
$\Omega_B = 60$~MHz and increases to 83\% for $\Omega_B =
120$~MHz. In return, micromotion sidebands, which are visible as
an additional structure in the spectrum calculated for $\Omega_B$
are smoothed by power broadening  for $2\Omega_B$. This dependence
on the Rabi frequencies implies that the use of the dark resonance
contrast for the reduction of the displacement $D_u$ has to be
made at constant laser power.

\section{Experimental results}
\label{exp} 
\begin{figure}
\begin{center}
\resizebox{0.75\textwidth}{!}{\epsfig{file=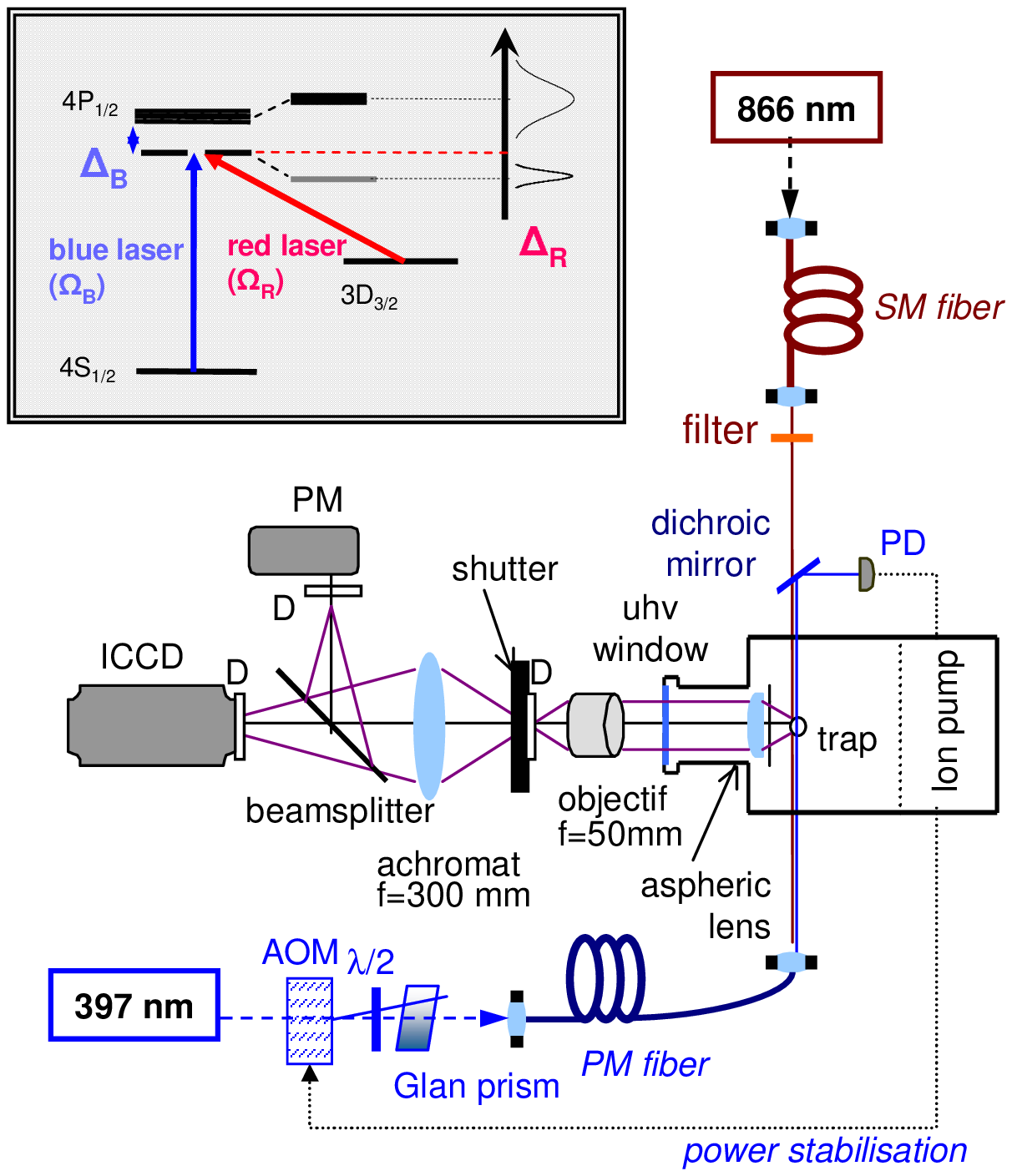}}
\end{center}
\caption{Experimental setup with counterpropagating laser beams.
The left-hand inset shows the first energy levels of a Ca$^+$-ion
dressed by the blue photons.} \label{fig:expset}
\end{figure}

We trap singly-ionized calcium in a miniature
cylindrical radiofrequency trap. The main laser-cooling transition
connects the 4$S_{1/2}$ ground state to the 4$P_{1/2}$-level at
397~nm (called the blue transition (B)), shown on the inset of
figure \ref{fig:expset}. This upper state has a lifetime of 7~ns
and a branching ratio to the metastable 3$D_{3/2}$-state of about
6\%. A closed laser-cooling cycle therefore requires a repumper
laser at 866~nm (red transition (R)) to empty the
3$D_{3/2}$-state. Our experimental set-up is based on a miniature
ring trap with an inner diameter of 1.4~mm. The trapping frequency
of $\Omega/2\pi \simeq$ 11.6~MHz with an amplitude of about
$V_{AC}$ = 300~V is applied to the ring, generating a
pseudo-potential well with a depth of a few $eV$, allowing the
confinement of ions for very long periods. Two tip electrodes in
the plane of the ring ($x, y$) and two larger mesh electrodes
along the ring axis ($z_1, z_2$), distant by approximately 5~mm
from the ring center, permit corrections of the potential by
application of direct voltages. The trapping characteristics of
this device are described in detail elsewhere \cite{champenois01}.

The experimental geometry is like in \cite{knoop04} and in figure
\ref{fig:expset}, where the case of counterpropagating laser beams
is depicted. The ultrahigh-vacuum vessel containing the miniature
trap allows the propagation of the laser beams along an
 axis having a 55~degree angle according to the trap's
$z$-axis. The ion's fluorescence is observed in the $xy$-plane of
the trap either spatially resolved  by an intensified CCD or as an
integrated  signal by a photomultiplier in photon-counting mode.
Laser beams are transported to the trap by single-mode fibers,
polarization-maintaining in the case of the blue laser. Laser
polarizations are modified by means of quarter- and
half-waveplates.

\begin{figure}
\begin{center}
  \resizebox{0.75\textwidth}{!}{ \epsfig{file=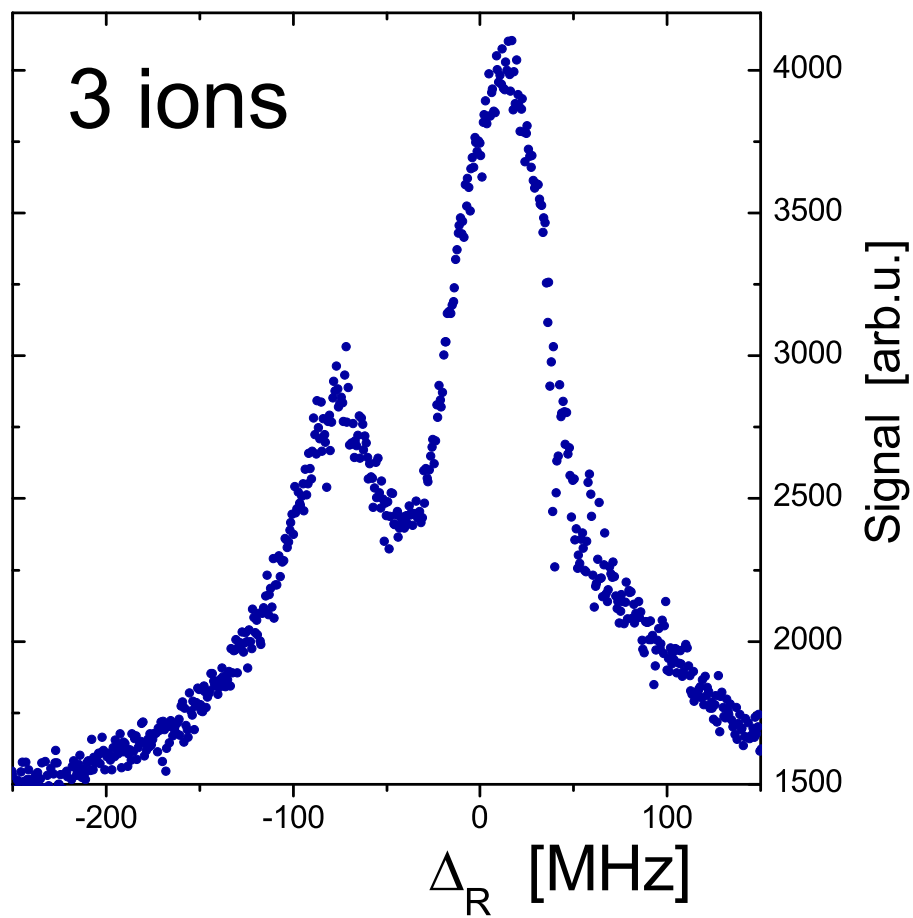}}
\end{center}
\caption{Dark resonance observed in the fluorescence of a 3-ion
cloud with copropagating laser beams.} \label{fig:3ions}
\end{figure}

Since dark resonances require identical detuning for both lasers,
their observation can be readily realized by keeping one laser
fixed, while probing the atomic resonance by scanning the other
laser. In order to  observe a full fluorescence spectrum without
losing the ion by Doppler laser heating,  the blue cooling laser
is fixed on the red-frequency side of the 4$S_{1/2}$-4$P_{1/2}$
transition (approximately at -~1 to -~2~$\Gamma_B$) and  the
frequency of the repumper laser at 866~nm is scanned. Throughout
the experiments presented in this section, the laser beam geometry
has been chosen to be counter-propagating, except for figure
\ref{fig:3ions}.

\begin{figure}
\begin{center}
   \resizebox{0.75\textwidth}{!}{ \epsfig{file=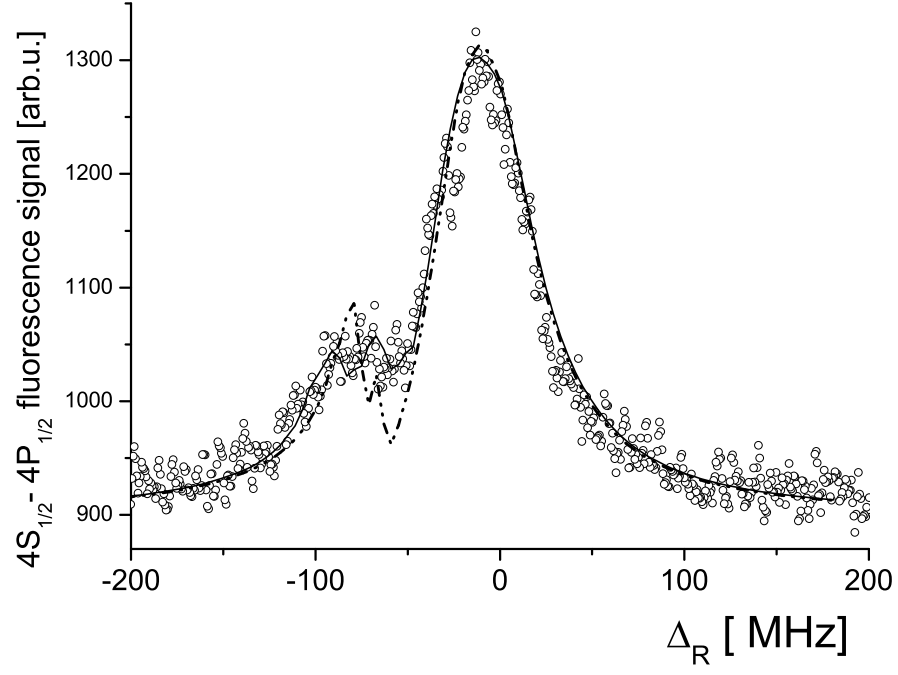}}
\end{center}
\caption{Experimental observation of a dark resonance ($\circ$),
adjusted by the numerical simulation with counter-propagating
laser beams (solid line). The dashed line is the numerical
simulation with an identical parameter set and with co-propagating
laser beams. ($\Omega_B$ = 60~MHz, $\Omega_R$ = 15~MHz, $\Delta_B$
= -50~MHz, $V_0$ = 1 m/s, and $D_u$ = 0.8~$\mu$m.)}
\label{fig:simul_exp}
\end{figure}

An experimentally recorded dark resonance signal as well as its
corresponding  numerical simulation are plotted in figure
\ref{fig:simul_exp}. The excellent agreement between experiment
and simulation allows to deduce the effective experimental
parameters, which in this case produce lower Rabi frequencies
($\Omega_B = 60$~MHz, $\Omega_R = 15$~MHz) than those used in
figures \ref{fig:simul} and \ref{fig:simul_rabi}. The dashed curve
on this graph is the simulation using exactly identical parameters
but changing the laser geometry to copropagating beams. The dark
resonance is then much more pronounced, confirming the dependance
of the signal with the laser beam geometry.

\subsection{Magnetic field}\label{sec:hanle}
\begin{figure}
\begin{center}
   \resizebox{0.75\textwidth}{!}{\epsfig{file=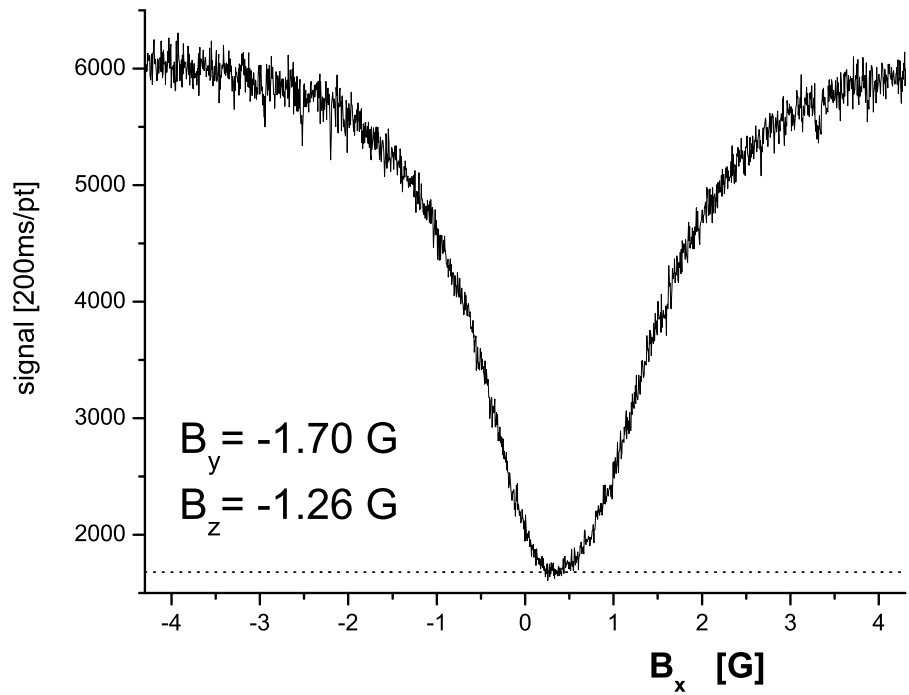}}
\end{center}
\caption{Fluorescence of a single ion versus the applied
perpendicular magnetic field in a Hanle type experiment  for the
determination of the magnetic field zero values. The red laser is
linearly polarized along the \textbf{x} direction. The dotted line
corresponds to the background light level. } \label{fig:hanle}
\end{figure}

 To make sure that laser polarization
effects do not play a role, the local magnetic field has to be
controlled in a rigorous way. We apply a  magnetic field
sufficient to define a quantization axis but small enough to limit
the splitting of the Zeeman sublevels  to values lower than the
natural width of the involved transition. In our experimental
setup, the local magnetic field is controlled by three pairs of
orthogonal Helmholtz coils. The residual
$\overrightarrow{B}$-field at the position of the ion has been
compensated making use of an optical pumping effect. Laser-cooling
is carried out on the 4$S_{1/2}$ - 4$P_{1/2}$ transition, while a
repumper on the 3$D_{3/2}$ - 4$P_{1/2}$ transition at 866~nm
avoids trapping of the ion in the metastable 3$D_{3/2}$ state. A
bias magnetic field $\vec B_0$ ($\approx$ 1~Gauss) is applied
perpendicularly to the propagation axis of the lasers, the
repumping laser being linearly polarized parallel to this field.
If the total magnetic field is parallel to $\vec B_0$, the ion is
pumped in the 3$D_{3/2}$, $m_J=\pm 3/2$ states, which are not
coupled to the laser, making the fluorescence disappear. Any
residual magnetic field perpendicular to $\vec B_0$ mixes these
dark states with the 3$D_{3/2}$, $m_J=\pm 1/2$ states and
fluorescence is then observed again. If the repumping laser
polarisation is purely linear, the observed fluorescence can be
reduced to zero when the applied magnetic field exactly
compensates the local field (see figure \ref{fig:hanle}). This
procedure is carried out for all three axis of the magnetic field,
changing the quantization axis and the corresponding laser
polarizations. The method, based on optical pumping in a dark
state, is independent on the linewidth and detuning of the lasers,
if we assume that the bias magnetic field is small enough to keep
the splitting of the Zeeman sublevels comparable to the excitation
linewidth of the lasers. As a consequence, this technique is not
sensitive to the motion of the ion and presents the advantage of
being useable for a single ion as well as for an
ion cloud. 

\subsection{Laser intensity}
\begin{figure}
\begin{center}
    \resizebox{0.75\textwidth}{!}{\epsfig{file=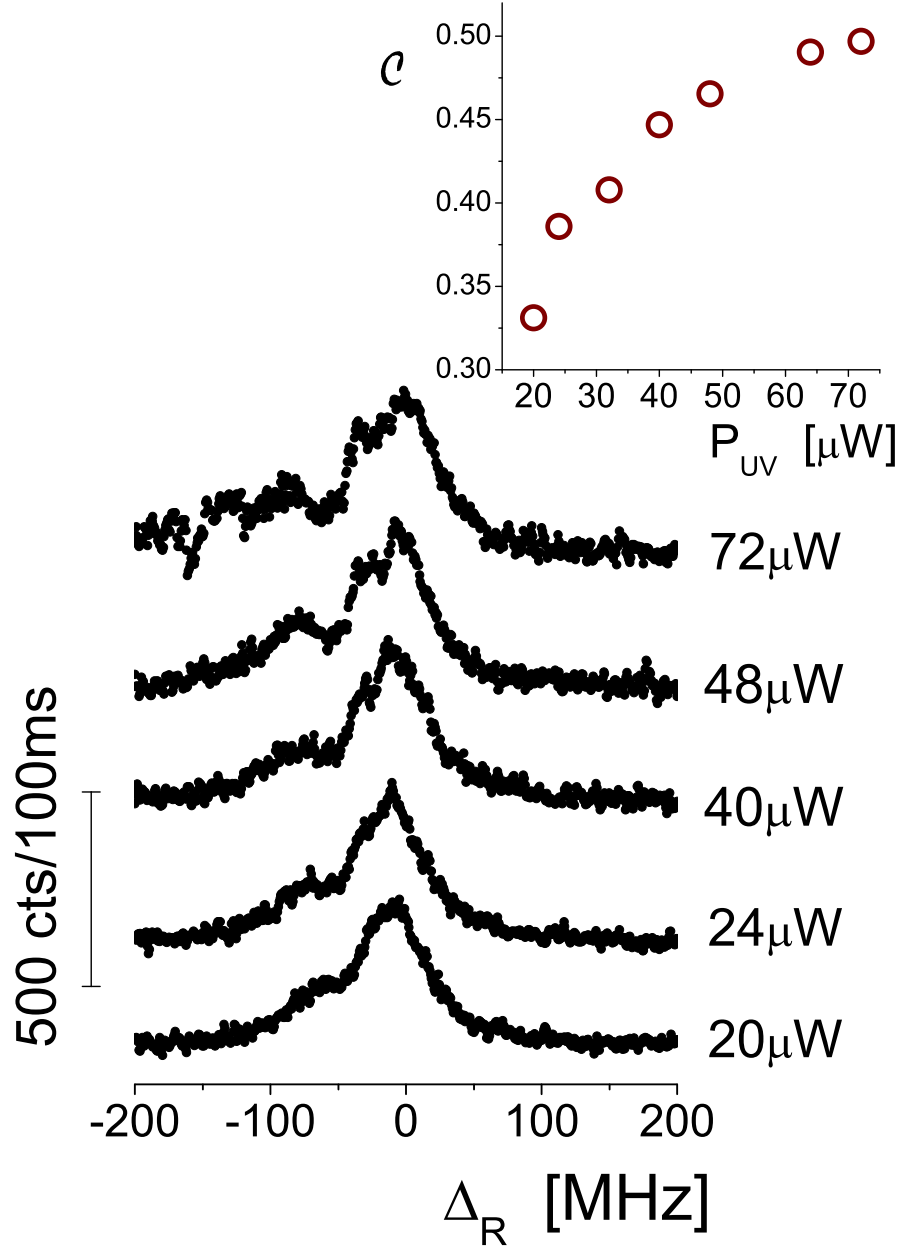}}
\end{center}
\caption{Fluorescence signal observed with counterpropagating
laser beams. The graph shows the influence of the laser power on
the level splitting, the offset of the curves is due to the
varying laser power. The applied blue laser powers are marked on
the right-hand side of the curves, where the lowest value
corresponds to about 2~$\Gamma_B$ while the highest power is
equivalent to 4~$\Gamma_B$. The inset shows the increasing
contrast of the fluorescence signal with higher laser power.}
\label{fig:varp397}
\end{figure}

As discussed above higher laser intensity at 397~nm splits the
coupled atomic levels to a greater extent and increases the
visibility of the Autler-Townes doublet. The five graphs in figure
\ref{fig:varp397} have been taken with increasing laser power,
starting from 20~$\mu$W,  which corresponds to a Rabi frequency of
about 2~$\Gamma_B$, as determined from numerical adjustments.  The
increase of the contrast with the applied blue laser power follows
a saturation curve as represented in the inset of the figure. For
laser intensities higher than 10~$\Gamma_B$ the transition is
largely saturated, and the maximum contrast is reached.

The influence of the red probe laser intensity has also been
checked while keeping the blue laser power fixed. As expected,
this only broadens the scanned atomic transition, without
increasing the contrast of the signal, as power broadening effects
prevail on the light shift effect.

\subsection{Micromotion}
\begin{figure}
\begin{center}
    \resizebox{0.75\textwidth}{!}{\epsfig{file=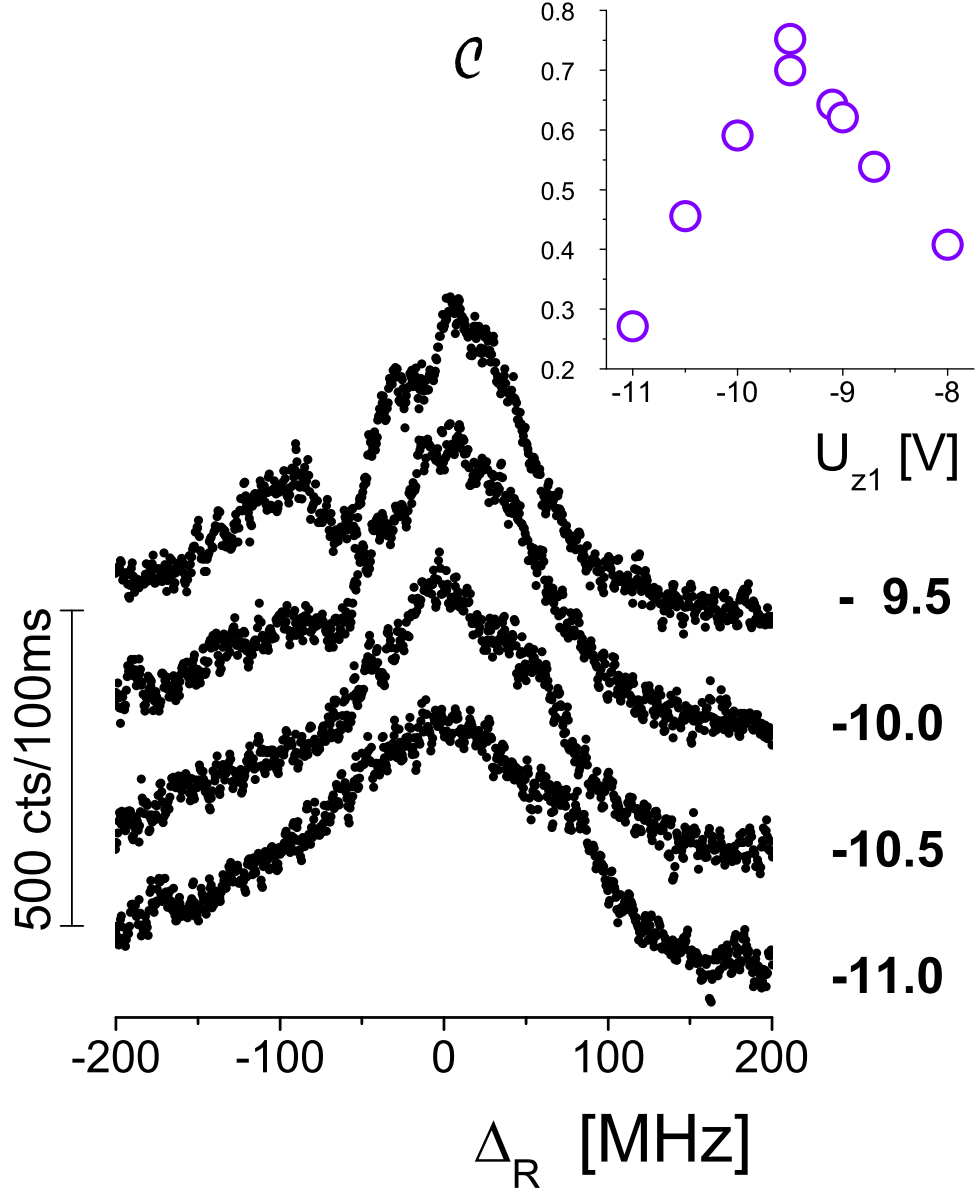}}
\end{center}
\caption{Influence of the micromotion on the observed
dark-resonance contrast $\mathcal C$ with counterpropagating laser
beams, laser intensity: $\Omega_B \approx 4\Gamma_B$, $\Omega_R
\approx 10\Gamma_R$. For better visibility, the graph only shows
selected curves which have been offset by a fixed value. The
compilation of the contrast data in the inset allows an
unambiguous determination of the best compensation parameter.}
\label{fig:varmicrom}
\end{figure}

For counterpropagating laser beams, dark resonances have merely be
observed in our experiment when the ion is  well localized, which
means that part of the excess micromotion has already been
eliminated by a rough compensation of the parasite potentials.
Simulations have shown the dependence of the contrast of the dark
resonance on the degree of localisation of the ion (see figure
\ref{fig:simul}). The smaller the ion's amplitude of motion, the
deeper the observed dark resonance. The curves plotted in figure
\ref{fig:varmicrom}  show the changing contrast in the observation
of the dark resonance, as a function of modified compensation
voltages applied on the correction electrodes. All curves have
been recorded at fixed blue and red laser intensity. Depending on
their distance from the center of the trap, a DC voltage applied
on one of the correction electrodes moves the ion by about 100 to
400~nm per Volt. Variations of a tenth of a Volt have an impact on
the micromotion and can be detected through the variation of the
dark resonance contrast. In addition, we have made sure, that the
applied static voltages do not modify the ion's position in the
laser beams. The inset of figure \ref{fig:varmicrom} shows the
contrast $\mathcal{C}$ of the observed signal versus the applied
compensation voltage and a maximum can be unambiguously
distinguished. We have verified that this maximum corresponds  to
a reduced spectral linewidth of the blue cooling transition. The
blue linewidth represents an estimate of the ion's kinetic energy
when approximated by a Doppler profile. It cannot be taken as an
absolute measure as it depends strongly on the applied (blue)
laser intensity, however it is a good relative indication of the
kinetic energy state of the ion, when the Doppler broadening is
larger than the power broadening. Compensation parameters have
been tested on all four correction electrodes, resulting in the
best localisation of a single ion within a precision of 10-40~nm,
which corresponds to the order of magnitude of the size of a
Doppler cooled wave packet.

\subsection{Laser beam geometry}
\label{sec:beamgeo} Equation (\ref{eq:doppler}) shows that the
dark resonance condition changes on the time scale of the
oscillation period of the ion, which is at least three orders of
magnitude smaller than the observation time. For fixed laser
detunings, the Doppler shifts fulfill the dark resonance condition
only for a fraction of the oscillation period, this fraction
decreases with increasing velocity amplitude and with increasing
wave vector difference $\vec{k_B}-\vec{k_R}$.  In the case of a
Ca$^+$ ion $|k_B|\,\approx\,|2\,k_R|$, we then expect the
sensitivity to ion oscillation three times higher for  the
counterpropagating configuration than for the copropagating one
(cf. figure \ref{fig:simul}).

This behaviour has been checked experimentally by using the
alternative beam geometry where both lasers enter the trap from
the same direction. In this configuration, dark resonances can be
observed in almost any case, even with a small ion cloud, which is
not possible in our experiment with the antiparallel geometry. As
an example, figure \ref{fig:3ions} shows the fluorescence spectrum
of three ions observed with copropagating laser beams.

\begin{figure}
\begin{center}
   \resizebox{0.75\textwidth}{!}{\epsfig{file=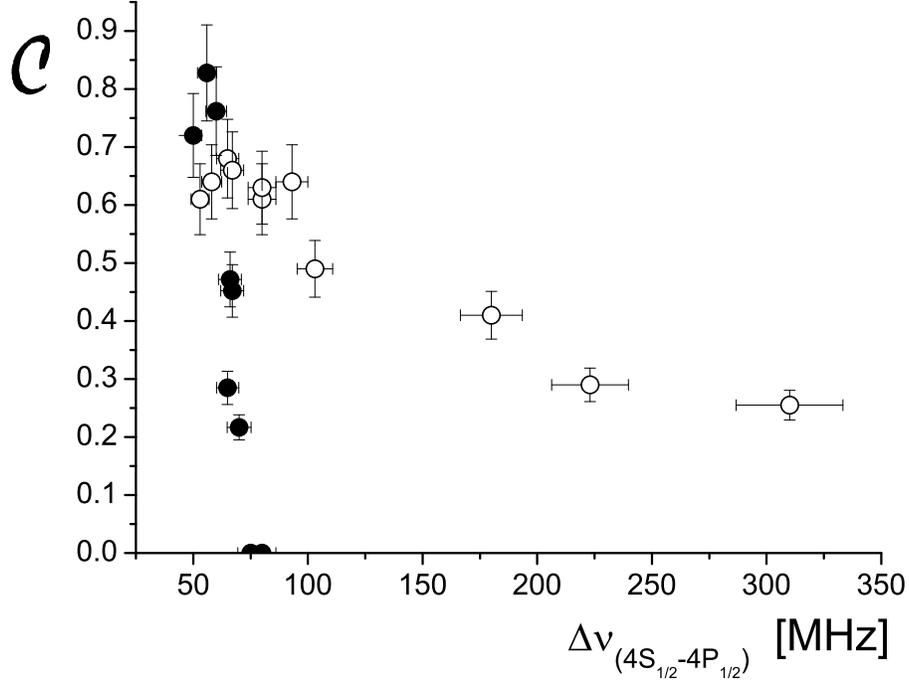}}
\end{center}
\caption{Comparison of  the dark resonance contrast for co-
($\circ$) and counter-propagating ($\bullet$) laser beam geometry.
In the graph the highest measured contrast is plotted as a
function of the observed transition linewidth on the
4S$_{1/2}$-4P$_{1/2}$ transition. The data for the copropagating
geometry has been taken with a single ion for linewidths below
about 100 MHz, and with very small ion clouds (inferior to 10
particles) for linewidths above. For this experiment, the blue
transition linewidths are due to micromotion and not to power
broadening. } \label{fig:DRsynth}
\end{figure}

To evidence the difference between co- and counterpropagating
laser beam geometry we have plotted the highest dark resonance
contrast $\mathcal{C}$ as a function of the experimental linewidth
of the 4$S_{1/2}$ - 4$P_{1/2}$ transition in figure
\ref{fig:DRsynth}. Again, the measured fluorescence  linewidth
gives a relative indication of the degree of localisation of the
trapped ion, as the smaller it is, the closer the ion is  to the
center of the
trap.  
The general tendency of figure \ref{fig:DRsynth} is clearly
visible: while dark resonances could not be observed at all for
the counterpropagating geometry unless a certain level of
localization has been reached, the copropagating geometry allows
the observation of dark resonances for much hotter ions and even
for small ion clouds. The highest contrast values achieved for
counterpropagating  laser beam geometry exceed 80\%, reflecting an
excellent localisation of the ion. In this beam configuration, the
steeper slope in the variation of the contrast values as depicted
in figure \ref{fig:DRsynth} can be used a sensitive tool to reveal
the degree of localization of the trapped ions with a high
sensitivity.

\section{Conclusion}
In this paper we have presented a novel approach to reduce the
micromotion of an ion confined in a miniature electrodynamic trap.
Only the lasers for cooling and detection of the ion are necessary
for the implementation of the proposed technique, in contrast to
other techniques, in particular the probing of the vibrational
sidebands which requires a highly stabilized laser to probe  a
narrow-linewidth transition. We use the contrast of the dark
resonances observed with two counterpropagating laser beams to
collect information about the motional state of the ion. Numerical
simulations and experiments have shown that this contrast is very
sensitive to the localisation of the ion, while a larger
micromotion amplitude smears out the coherent population trapping
effect. Observation of fluorescence is made on time bins which are
long compared to the inverse motional frequencies of the ion. On
this timescale, the oscillation amplitudes along the three
directions are coupled in the trap.

Comparison with other compensation techniques is not
straightforward as the different methods produce various signal
levels and use different sets of experimental parameters. In any
case, the maximum fluorescence signal has to be monitored as a
function of the compensation voltages, to make sure that a
decrease in the observed signal height is not due to the fact that
the ion has been pushed out of the laser beams.  The spatial
displacement of the confined particle with lowered
pseudo-potential well depth requires the use of a detector with
spatial resolution and can  only be applied to reduce effects
which appear in the plane of observation. The exploration of the
linewidth on the 4$S_{1/2}$ - 4$P_{1/2}$-transition is an easy
method to roughly minimize excess micromotion, keeping in mind
that the linewidth depends also on the applied laser intensities
and frequencies which may modify the efficiency of laser cooling.
Since excess micromotion may also deform the excitation spectrum
of the ion, it is  important to  record the entire frequency
response at fixed laser intensities. The fluorescence correlation
method which shows the modulation of the ion's fluorescence with
the micromotion frequency $\Omega$ gives best signal-to-noise
ratios in the low intensity limit but requires an integration time
of at least a couple of seconds to accumulate sufficient signal in
each time channel of a time-to-amplitude converter
\cite{berkeland98}.

Due to the wide panel of recorded signals a quantitative
comparison of the different compensation methods cannot be made in
a precise way. As a matter of fact, best results for the reduction
of excess micromotion are obtained by using at least two different
methods. Measurements along multiple laser axes have to be
performed to assure a complete extinction of excess micromotion
for all projections in space. We have shown that even in the case
where the blue laser linewidth is not modified, dark resonances
can still be observed. It is thus possible to use this means of
micromotion reduction, when the rf correlation technique and the
observation of the blue transition linewidth fail. The main
advantage of the dark resonance technique is nevertheless its
simplicity and the absence of need for any other recording or
excitation devices than those used for fluorescence observation.

In summary, we propose a complementary method to reduce excess
micromotion of a single trapped ion making use only of the
experimental means necessary to detect the ion's fluorescence. In
a more general way, this technique can be applied to any
three-level ion having a $\Lambda$ energy scheme, the highest
sensitivity being reached for $k_B = k_R$.



\newpage

\end{document}